\documentclass{bioinfo}
\copyrightyear{2013}
\pubyear{2013}

\usepackage{subfigure}
\usepackage{amsmath}
\usepackage{amssymb}
\RequirePackage{natbib}
\usepackage{algorithm}
\usepackage{algorithmic}
\usepackage{xspace}
\usepackage{tabularx}

\usepackage{url}
\usepackage{color}
\definecolor{red}{rgb}{0.8,0,0}
\newcommand\comment[1]{{\sf\footnotesize\textcolor{red}{$[[$#1$]]$}}}

\newcommand{\x}{{\boldsymbol{x}}}

\newcommand{\y}{{\boldsymbol{\pi}}}

\newcommand{\valpha}{{\boldsymbol{\alpha}}}

\newcommand{\vatilt}{\boldsymbol{\bar{a}}}
\newcommand{\btilt}{\bar{b}}

\newcommand{\va}{{\boldsymbol{a}}}

\newcommand{\EER}{$\hat{L}$\xspace}

\newcommand{\RR}{\mathbb{R}}

\newcommand{\argmax}{\mathop{\mathrm{argmax}}}
\newcommand{\argmin}{\mathop{\mathrm{argmin}}}

\DeclareMathOperator{\w}{\boldsymbol{\theta}}

\begin{document}

\firstpage{1}

\title[Transcript inference with mTim]{mTim: rapid and accurate transcript reconstruction from RNA-Seq data}
\author[G.Zeller \textit{et~al.}]{Georg Zeller\,$^{\ast,1,2,4}$, Nico G\"ornitz\,$^{\ast,3}$, Andr\'{e} Kahles\,$^{1,5}$, Jonas Behr\,$^{1,5}$, Pramod Mudrakarta\,$^{1}$, S\"oren Sonnenburg\,$^{6}$, and Gunnar R\"atsch\,$^5$\footnote{to whom correspondence should be addressed}
}
\address{
$^{1}$Friedrich Miescher Laboratory, Max Planck Society, 72076 T\"ubingen, Germany\\
$^{2}$Department of Molecular Biology, Max Planck Society, 72076 T\"ubingen, Germany\\
$^{3}$Machine Learning Group, Technical University Berlin, 10587 Berlin, Germany\\
$^{4}$Structural and Computational Biology Unit, EMBL, 69117 Heidelberg, Germany\\
$^{5}$Computational Biology Center, Sloan-Kettering Institute, NY 10065, USA\\
$^{6}$TomTom, An den Treptowers 1, 12435 Berlin, Germany \\
$^\ast$Authors contributed equally.
}

\history{Received on March 30th, 2013; revised on XXXXX; accepted on XXXXX}
\editor{Associate Editor: XXXXXXX}

\maketitle

\begin{abstract}

\section{Motivation:}
Recent advances in high-throughput cDNA sequencing (RNA-Seq) technology have revolutionized transcriptome studies. A major motivation for RNA-Seq is to map the structure of expressed transcripts at nucleotide resolution. With accurate computational tools for transcript reconstruction, this technology may also become useful for genome (re-)annotation, which has mostly relied on de novo gene finding where gene structures are primarily inferred from the genome sequence.

\section{Results:}
We developed a machine-learning method, called mTim
(\textbf{m}argin-based \textbf{t}ranscript \textbf{i}nference \textbf{m}ethod) for transcript reconstruction from RNA-Seq read alignments that is based on discriminatively trained hidden Markov support vector machines. In addition to features derived from read alignments, it utilizes characteristic genomic sequences, e.g. around splice sites, to improve transcript predictions. mTim inferred transcripts that were highly accurate and relatively robust to alignment errors in comparison to those from Cufflinks, a widely used transcript assembly method.
\section{Availability:}
Source code in Matlab/C is available from \url{https://github.com/nicococo/mTIM}. An mTim predictor is also provided as part of Oqtans, a Galaxy-based RNA-Seq analysis pipeline (\url{http://oqtans.org/}). 

\section{Contact:} \href{ratschg@mskcc.org}{ratschg@mskcc.org}
\end{abstract}

\section{Introduction}

High-throughput sequencing technology applied to cellular mRNA (RNA-Seq) has revolutionized transcriptome studies~\cite[among many others]{AMortazavi2008, JMarioni2008, ZWang2009}. In contrast to microarray platforms, which it has replaced in many applications, RNA-Seq can not only be used to accurately quantify known transcripts, but also to reveal the precise structure of transcripts at single-nucleotide resolution. RNA-Seq based transcript reconstruction has therefore become a valuable tool for the completion of genome annotations~\cite[for instance]{ARoberts2011, XGan2012} and further enabled subsequent  analyses of differentially expressed genes~\cite{SAnders2010}, transcript isoforms~\cite{RBohnert2010, JBehr2013} 
and exons~\cite{SAnders2012},
all of which generally rely on correctly inferred transcript inventories. De novo transcript reconstruction is thus a pivotal step in the analysis of RNA-Seq data. 

There are two conceptually different strategies to approach this problem: one can either assemble transcripts directly from RNA-Seq reads using methodology that originated from genome assembly approaches~\cite{MGrabherr2010, GRobertson2010, MSchulz2012}. Alternatively, the problem can be decomposed into two steps: RNA-Seq reads are first aligned to the genome of origin followed by the actual transcript reconstruction on the basis of these alignments. While the first, assembly-based strategy does not require a high-quality genome sequence and is thus applicable to non-model organisms, it is arguably addressing a more difficult problem than the latter, mapping-based approach. Consequently, transcripts, in particular ones with low expression, may be more accurately reconstructed by methods implementing the mapping-based approach~\cite{CTrapnell2010, MGuttman2010, AMezlini2012} (see also \cite{MGrabherr2010, MSchulz2012} for a comparison). The performance of mapping-based methods however strongly depends on the quality of the RNA-Seq read alignments. Considerable attention has therefore been payed to solve the problem of correctly aligning RNA fragments across splice junctions~\cite{FDeBona2008, CTrapnell2009, GJean2010}. 

Following the mapping-based paradigm, we developed a novel machine learning-based method, which we call mTim: \textbf{m}argin-based \textbf{t}ranscript \textbf{i}nference \textbf{m}ethod. In contrast to algorithmic transcript assembly \cite{CTrapnell2010, MGuttman2010}, we formalize the problem as a supervised label sequence learning task and apply state-of-the-art techniques, namely Hidden Markov support vector machines (HM-SVMs)~\cite{YAltun2003, ITsochantaridis2006, GRatsch2007NIPS, GZeller2008PSB}. This way of approaching the problem is similar to recently developed gene finders \cite{GSchweikert2009, SGross2007}, and mTim is indeed a hybrid method that can utilize both, RNA-Seq read alignments and characteristic features of the genome sequence, e.g. around splice sites \cite{SSonnenburg2007}. However, mTim's emphasis is on inference from aligned RNA-Seq reads, and its model is only augmented by a few genic sequence motif sensors \cite{GSchweikert2009}, which can moreover be disabled. We thus make weak assumptions, if any, about the inferred transcripts: importantly, we do not model protein-coding sequences (CDS) and are thus able to predict noncoding transcripts as well as coding ones with similar expression.


\begin{methods}
\section{Methods}

\subsection{Transcript reconstruction}

The task of reconstructing the exon-intron structure of expressed genes can be converted into a label sequence learning problem, where we attempt to label each nucleotide in the genome as either intergenic, exonic or intronic. Our prior knowledge about what constitutes a valid gene structure is incorporated into a state model to restrict the
space of possible labelings to valid ones.

\subsubsection{The state model used by mTim}

Starting from a naive state model that would consist of a single state for each of
the atomic labels, exonic, intronic, and intergenic, we extended it as follows
(see Figure~\ref{fig:model}): first, we devised a strand-specific model. Second, we created expression-dependent submodels. This allows us
to maintain several parameter sets, each of which is optimized for
transcripts with a certain read support. Due to non-uniform read
coverage along transcripts, transitions between expression levels
proved useful in practice. Finally, the simple model was extended by states that mark segment boundaries (e.g. when transitioning from exon to intron), as this facilitates boundary recognition from features such as spliced reads (Fig.~\ref{fig:model}).

\begin{figure}[!tpb]
\vspace*{-2ex}
\centerline{\includegraphics[width=0.9\columnwidth]{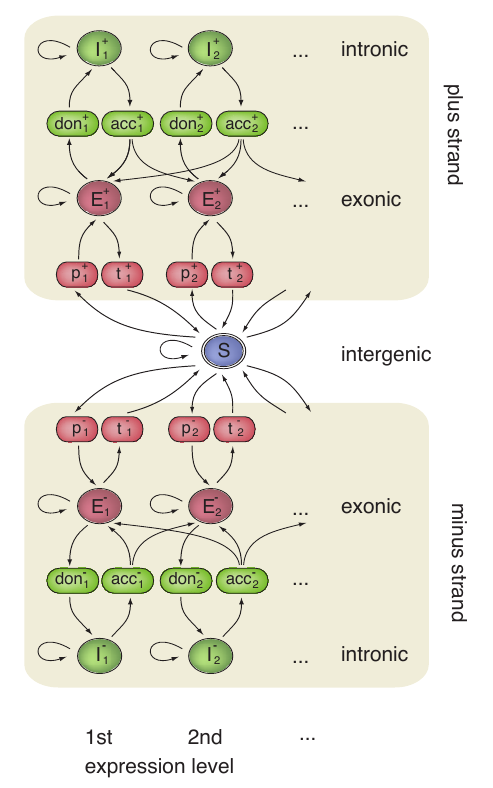}}
\vspace*{-2ex}
\caption{State model used by mTim. Ovals correspond to states,
  whereas arrows indicate allowed transitions. The correspondence
  between states and atomic labels (exon, \texttt{E}, intron, 
  \texttt{I}, and intergenic, \texttt{S}; see Methods) is
  color-coded. The first and last nucleotide of introns and transcripts were modelled with particular care: The former are associated with splice
  site signals at exon-intron junctions (states denoted 
  \texttt{acc} and \texttt{don}), whereas the latter correspond
  to transcript start and end (denoted \texttt{p} and \texttt{t}, respectively). 
  The model is strand-specific (superscripts \texttt{+} and \texttt{-} in state labels)
  and consists of expression-specific submodels (columns and
  subscripts, see also label at bottom) which allows us to optimize
  different parameter sets depending on the read
  support.}\label{fig:model}
\end{figure}

\subsubsection{Feature derivation and training data}

The inference of transcript structures is based on sequences of observations or features derived from RNA-Seq read alignments and
predicted splice sites. Specifically, we derive the following position-wise features from RNA-Seq alignments:
\begin{itemize}
\item number of reads aligned at the given position, indicating an
  exon.
\item a gradient of the read coverage; high absolute values correspond to sharp in- or decreases in coverage typical of the 
  start and end of exonic regions, respectively
\item number of reads that are spliced over the given position
  (strand-specific), thus indicating an intronic position.
\item number of spliced reads supporting a donor splice site at the
  given position (strand-specific).
\item number of spliced reads supporting an acceptor splice site at the given position
  (strand-specific).
\item number of paired-read alignments for which the insert spanned the given position (only used if
  read pair information is available, strand-specific), an indicator
  of transcript connectivity.
\end{itemize}

Additionally, we derive features from the genome sequence around a given position such as strand-specific donor and acceptor splice site prediction.

As a ground truth for guiding the supervised training process,
annotated gene models with a portion of the surrounding intergenic region are excised and converted into label sequences by assigning one of the above atomic labels to each nucleotide (see color coding in Figure~\ref{fig:model}).  In the presence of alternative transcripts, this labeling was based on a single isoform (the one that was best supported by RNA-Seq reads), and additionally a mask of alternative transcript regions was generated to avoid that learning the correct alternatives is penalized during training.

\subsubsection{Training and optimization}

Label sequence learning problems are often addressed using Hidden Markov Models (HMMs) \cite{RDurbin1998}. However, discriminative learning algorithms have recently been developed that combine the versatility of HMMs with the advantages of discriminative training, and these include hidden Markov support vector machines (HMSVMs)\cite{YAltun2003, ITsochantaridis2006, GRatsch2007NIPS}. Inspired by support vector machines~\cite{VVapnik1995, BScholkopf2002, ABenHur2008}, the HMSVM training problem amounts to maximizing the margin of separation between the correct and any wrong label sequence.

Formally, training an HMSVM involves learning a function
\[
f:X\rightarrow\mathcal{S}^{\star}
\]
that yields a label sequence (or simply a path)
$\pi\in\mathcal{S}^{\star}$ given the corresponding sequence of
observations ${\bf x}\in X$ (an $m\times t$ matrix of $m$ different
features), both of length $t$, where $\mathcal{S}^{\star}$ denotes the
Kleene closure of the state set (see Figure~\ref{fig:model}). This is done
indirectly via a $\boldsymbol{\theta}$-parametrized discriminant
function
\[
F_{\boldsymbol{\theta}}:X\times\mathcal{S}^{\star}\rightarrow\RR
\]
that assigns a real-valued score to a pair of observation and state
sequence \cite{YAltun2003}. Once $F$ is known, $f$ can be obtained
as
\[
f({\bf x}) = \argmax\limits_{\boldsymbol{\pi}\in\mathcal{S}^{\star}}
 F_{\boldsymbol{\theta}}({\bf x},\boldsymbol{\pi}).
\]
In our case, $F$ satisfies the Markov property and can hence be
efficiently decoded using the Viterbi algorithm
\cite{RDurbin1998}.

The discriminant function is essentially a linear combination of
feature scoring functions $g_{j,k}$ (piece-wise linear transformations
of feature values into real-valued scores, see
\cite{GRatsch2007PlosCompBiol} for details) and transition scores
$\phi$.
\[
F_{\boldsymbol{\theta}}({\bf x},\boldsymbol{\pi})=
\sum\limits_{p=1}^{t}
\big( \sum\limits _{j=1}^{m}
\sum\limits_{k\in\mathcal{S}}[[\pi_{p}=k]]
\: g_{j,k}(x_{j,p}) \big)
+ \phi(\pi_{p-1},\pi_{p})
\]
where {[}{[}.]] denotes the indicator function. The parametrization of
the feature scoring functions $g_{j,k}$, together with the transition
scores $\phi$ constitutes the parametrization of the model denoted by
$\boldsymbol{\theta}$.

Let $n$ be the number of training examples $({\bf
  x}^{(i)},\boldsymbol{\pi}^{(i)}),\: i=1,\ldots,n$. Following the
discriminative learning paradigm, we want to enforce a large margin of
separation between the score of the correct path
$\boldsymbol{\pi}^{(i)}$ and \emph{any} other wrong path
$\overline{\boldsymbol{\pi}}\neq\boldsymbol{\pi}^{(i)}$, i.e.,
\[
F_{\theta}({\bf x}^{(i)},\boldsymbol{\pi}^{(i)})
 - F_{\theta}({\bf x}^{(i)},\overline{\boldsymbol{\pi}})\;
 \gg\;0
 \qquad\forall\overline{\boldsymbol{\pi}}\neq\boldsymbol{\pi}^{(i)}
 \quad\forall i=1,\ldots,n
\]
To achieve this, we solve the following optimization problem: 
\[
\min\limits_{\boldsymbol{\theta},\:\boldsymbol{\xi}
 \geq{\bf 0}}\quad\frac{1}{n}\sum\limits_{i=1}^{n} 
 \xi^{(i)}+C\:\Omega(\boldsymbol{\theta})
\]
\begin{align}\label{eq:sosvm}
 \text{s.t.}\quad F_{\theta}({\bf x}^{(i)},\boldsymbol{\pi}^{(i)})
 - F_{\theta}({\bf x}^{(i)},\overline{\boldsymbol{\pi}})\;
 \geq\;\Delta(\y^{(i)},\overline{\y})-\xi^{(i)}
 \quad\forall\overline{\boldsymbol{\pi}}\neq\boldsymbol{\pi}^{(i)}
 \:\forall i
\end{align}
where $\Omega$ is a regularization term to restrict model complexity
(see \cite{GZeller2008GenomeRes, GZeller2008PSB} for details), whose weight is
adjusted through the hyper-parameter $C$. So-called slack variables
$\xi^{(i)}$ implement a soft-margin \cite{CCortes1995} allowing for
some prediction errors on the training set.

\paragraph{A Bundle Method for Efficient Optimization}
A common approach to obtain a solution to \eqref{eq:sosvm} is to use so-called cutting plane or column generation methods. These accumulate growing subsets (working sets) of all possible structures and solve restricted optimization problems \cite{ITsochantaridis2006}. However, these techniques often converge slowly. Moreover, the size of the restricted optimization problems grows steadily and solving them becomes more expensive in each iteration. Simple gradient descent or second order methods can not be directly applied as alternatives, because the above objective function 
is continuous but non-smooth. Our approach is instead based on bundle methods for regularized risk minimization \cite{SmoVisLe08, TeoVisSmoLe10, Do10}. In order to achieve fast convergence, we use a variant of these methods adapted to structured output learning.

We consider the objective function $J(\w) = \Omega_{p,\gamma}(\w) + L(\w)$, where 
$$
L(\w): = C\sum_{i=1}^n \ell( \max_{\bar{\y}\in\Upsilon}\ \{ \langle\w, \Psi(\x_i,\bar{\y})\rangle + \Delta(\y_i,\bar{\y}) \} - \langle\w, \Psi(\x_i,\y_i)\rangle ) \, .
$$
Direct optimization of $J$ is very expensive as computing $L$ involves computing the maximum over the output space. Hence, we propose to optimize an estimate of the empirical loss \EER$(\w)$, which can be computed efficiently. 
We define the estimated empirical loss \EER$(\w)$ as 
\vspace*{-2ex}
$$
\hat{L}(\w) := C\sum_{i=1}^N \ell\left( \max_{(\Psi,\Delta)\in\Gamma_i}\ \{\langle\w, \Psi\rangle + \Delta\} - \langle\w, \Psi(\x_i,\y_i)\rangle \right).
$$
Accordingly, we define the estimated objective function as $\hat{J}(\w) = \Omega_{p,\gamma}(\w) +
\hat{L}(\w)$. It is easy to verify that $J(\w) \geq \hat{J}(\w)$.  $\Gamma_i$ is a set of pairs
$(\Psi(\x_i,\y), \Delta(\y_i,\y))$ defined by a suitably chosen, growing subset of $\Upsilon$, such that $\hat{L}(\w)\to L(\w)$ (cf.\ 
Algorithm~\ref{alg-gd}).

In general, bundle methods are extensions of cutting plane methods that use a prox-function to stabilize the solution of the approximated function. In the framework of regularized risk minimization, a natural prox-function is given by the regularizer. We apply this approach to the objective $\hat{J}(\w)$ and solve
\vspace*{-2ex}
\begin{align}
  \label{eq:bmrm}
  \min_{\w} \Omega_{p,\gamma}(\w) + \max_{i \in I} \{\langle\va_i, \w\rangle + b_i   \}
\end{align}
where the set of cutting planes $\va_i$, $b_i$ lower bound \EER. As proposed in
\cite{Do10,TeoVisSmoLe10}, we use a set $I$ of limited size. 
Moreover, we calculate an aggregation cutting plane $\vatilt$, $\btilt$ that lower bounds the
estimated empirical loss \EER. To be able to solve the primal
optimization problem in \eqref{eq:bmrm} in the dual space as proposed
by \cite{Do10,TeoVisSmoLe10}, we adopt an elegant strategy described in \cite{Do10} to obtain the
aggregated cutting plane $(\vatilt',\btilt')$ using the dual solution $\valpha$ of \eqref{eq:bmrm}: 
\begin{equation} 
\vatilt' = \sum_{i\in I} \alpha_j \va_i\qquad\mbox{and}\qquad \btilt' = \sum_{i\in I} \alpha_i b_i. \label{eq:agg}
\end{equation}
\vspace*{-1ex}
The following two formulations reach the same minimum when optimized with respect to $\w$:
$$
\min_{\w\in\mathcal{H}}\ \left\{\Omega_{p,\gamma}(\w) + \max_{i \in I} \langle\va_i, \w\rangle + b_i\right\} = \min_{\w\in\mathcal{H}}\ \{ \Omega_{p,\gamma}(\w) +
\langle\vatilt', \w\rangle + \btilt'\}.
$$
This new aggregated plane can be used as an additional cutting plane in the next iteration step. We therefore have a monotonically increasing lower bound on the estimated empirical loss and can remove previously generated cutting planes without compromising convergence (see \cite{Do10} for details).

The algorithm is able to handle any (non-)smooth convex loss function $\ell$, since only the subgradient needs to be computed. This can be done efficiently for the hinge-loss, squared hinge-loss, Huber-loss, and logistic-loss.

The resulting optimization algorithm is outlined in Algorithm~\ref{alg-gd}.

\begin{algorithm}
\caption{Bundle Methods for Structured Output Algorithm} \label{alg-gd}
\begin{algorithmic} [1]
\STATE $S\geq 1$: size of the bundle set 
\STATE $\tau>0$: linesearch trade-off (cf.\ \cite{FraSon08} for details)
\STATE $\w^{(1)} = \w_p$
\STATE $k=1$ and $\vatilt={\bf 0}, \bar{b}=0, \Gamma_i=\emptyset \quad \forall i$
\REPEAT
\FOR{$i = 1,..,n$} 

    \STATE $\y^\ast = \argmax_{\y \in \Upsilon} \{\langle\w^{(k)},\Psi(\x_i,\y)\rangle + \Delta(\y_i,\y) \}$ 
    \STATE $\text{MMV}_\text{true} := \max_{\y \in \Upsilon}\ \{\langle \w, \Psi(\x_i,\y)\rangle + \Delta(\y_i,\y)\}$ 
    \STATE $\text{MMV}_\text{est} := \max_{(\Psi,\Delta)\in\Gamma_i} \langle \w,\Psi \rangle + \Delta $ 
    
    \IF{$\displaystyle\ell\left(\text{MMV}_\text{true}\right) > \ell\left(\text{MMV}_\text{est}\right)$} 
        \STATE $\Gamma_i = \Gamma_i \cup (\Psi(\x_i,\y^\ast), \Delta(\y_i,\y^\ast))$
    \ENDIF
    
    \STATE Compute $\va_k \in \partial_{\w} \hat{L}(\w^{(k)})$ 
    \STATE Compute $b_k = \hat{L}(\w^{(k)}) - \langle\w^{(k)},\va_k\rangle$
    \STATE Define $Z_k(\w) :=  \max_{(k-S)_+ < i \leq k} \{\langle\va_i, \w\rangle + b_i\}$
    \STATE $\w^{*} = \displaystyle\argmin_{\w\in\mathcal{H}}\ \left\{\Omega_{p,\gamma}(\w)+ \max\left( Z_k(\w), \langle\vatilt, \w\rangle + \btilt \right)\right\}$
    \STATE Update $\vatilt$, $\btilt$ according to \eqref{eq:agg} 
    \STATE $\eta^{*} = \argmin_{\eta \in \Re} \hat{J}(\w^{*} + \eta(\w^{*} - \w^{(k)}))$
    \STATE $\w^{(k+1)} = (1-\tau) \w^{*} + \tau \eta^{*} (\w^{*} - \w^{(k)})$
    \STATE $k=k+1$
\ENDFOR
\UNTIL{$L(\w^{(k)})- \hat{L}(\w^{(k)})\leq \epsilon$ and $J(\w^{(k)}) - J_k(\w^{(k)}) \leq \epsilon$}
\end{algorithmic}
\end{algorithm}

\subsection{Data preparation and feature generation}

\subsubsection{RNA-Seq alignments}
For the following computational experiments we used RNA-Seq data from well-studied model organism for which high-quality annotations exist, because these can not only be used for training, but also to assess the accuracy of the inferred transcripts.

We aligned RNA-Seq reads to the genome using the splice-aware alignment tool PalMapper \cite{GJean2010}

\subsubsection{Alignment filtering}
Primary RNA-Seq alignments were filtered with the goal to reduce the number of alignment errors. To this end, we used a small subset of annotated introns to define an optimal choice of parameters for filtering criteria such as maximal number of edit operations (mismatches, insertions, and deletions), minimal length of the shortest aligned segment in a spliced alignment, and the minimal number of alignments supporting an intron. The chosen filter settings maximize the F-Score (harmonic mean of precision and recall) between the annotation set and the introns contained in the filtered alignments. 

\subsubsection{Splice site prediction}
Donor and acceptor splice sites were predicted from the genome sequence following a published protocol \cite{SSonnenburg2007}. In summary, this method cuts out genomic sequences around all potential splice donor and acceptor site (exhibiting the two-nucleotide consensus sequence) and applies SVM classifiers with string kernels to recognize annotated splice sites. Trained classifiers are subsequently used to generate whole-genome predictions which were subsequently transformed into probabilistic confidence values~\cite{SSonnenburg2007}.


\subsubsection{Feature and label generation from RNA-Seq alignments}

From the RNA-seq read alignments we then generated the above-listed coverage and splice-site features and derived a label sequence from the corresponding gene annotations (see above for details).
 
\subsection{Design of computational experiments}

To be able to assess the impact of alignment quality on subsequent transcript inference, we used unfiltered alignments in a first set of experiments and subsequently repeated these using filtered RNA-seq alignments as input to assess the improvement of transcript inference with improved alignment quality.

To generate transcript models from these read alignments, the mTim pipeline proceeds through the following steps:

\begin{enumerate}
\item Definition of genome chunks; importantly, chunks are defined based on read coverage only without using any annotation information.
\item Partitioning genome chunks into subsets for cross validation.
\item Training on chunks from the training set using known (annotated)
  gene models as ground truth.
\item Application of the trained mTim models to predict transcript structures on test chunks.
\end{enumerate}

Using cross-validation, we obtain unbiased estimates of mTim's transcript reconstruction accuracy for data it had not seen during training.

To compare mTim's prediction to the state of the art in alignment-guided transcript inference, we also applied Cufflinks with default parameter settings to the same unfiltered and filtered RNA-seq alignment data.




\end{methods}


\section{Results and Discussion}


\begin{figure*}[bt]
\parbox{\textwidth}{
  \parbox{1.395\columnwidth}{
    \includegraphics[width=1.42\columnwidth]{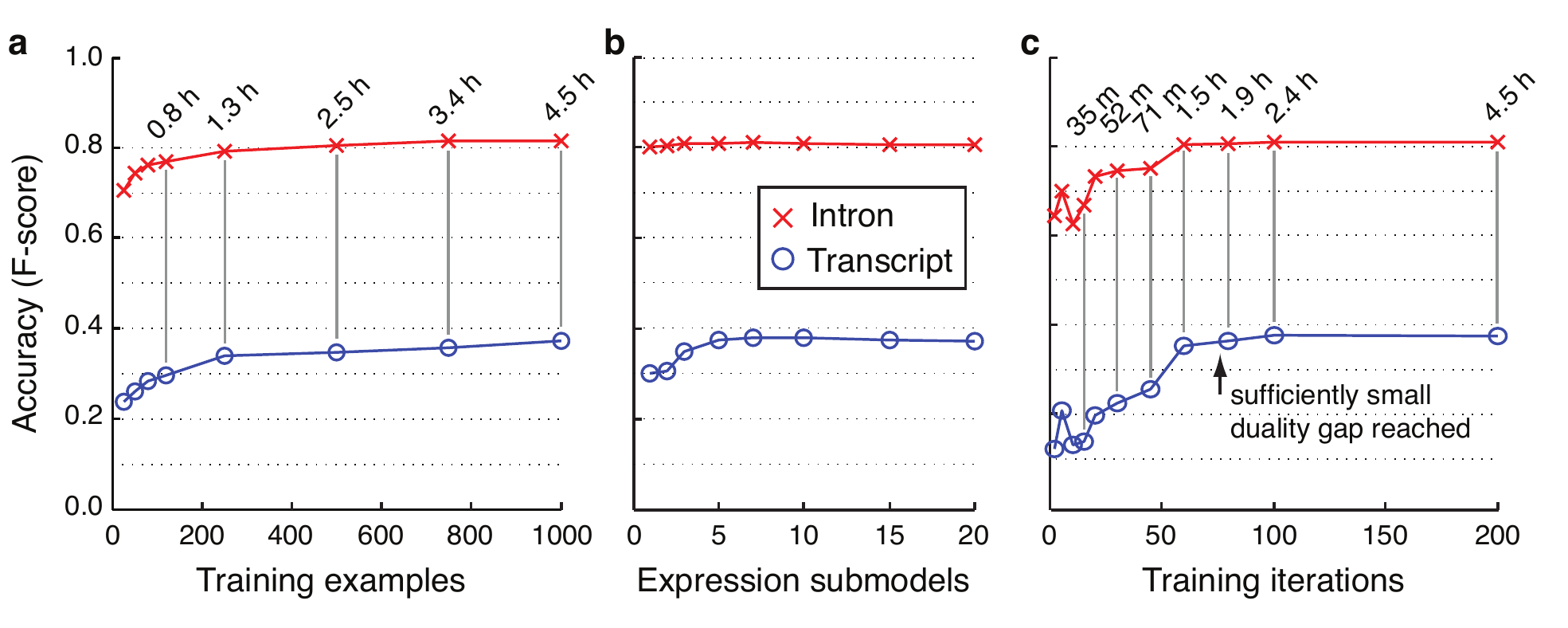}
  } \hfill
    \parbox{0.6\columnwidth}{
    \caption{Opimizing mTim's performance. 
    \textbf{(a)} The HM-SVM learning algorithm
    utilized training data efficiently and accuracy quickly reached a plateau.
    \textbf{(b)} Expression-specific submodels
    (see Fig.~\ref{fig:model}
    and Methods)
    improve reconstruction of complete transcripts.
    \textbf{(c)} Accuracy as a function of the number of training
    iterations (using 1000 examples). The duality gap was
    sufficiently small for termiantion after 78 iterations. All
    results were obtained using unfiltered RNA-Seq 
    alignments for \textit{C.\ elegans}. Empirical execution times 
    in (a) and (c) were averaged across three HM-SVM trainings. 
    }
  }
}
\vspace*{-2ex}
\label{fig:train}
\end{figure*}

To evaluate its performance, we applied mTim to RNA-Seq data from model species. We chose three organisms, \textit{Chaenorhabditis elegans} (nematode worm), \textit{Arabidopsis thaliana} (thale cress) and \textit{Drosophila melanogaster} (fruit fly), whose genomes and transcriptomes have been extensively characterized~\cite{MGerstein2010, XGan2012}, 
making it possible to use annotated gene models as a ground truth for evaluating the quality of transcripts reconstructed from RNA-Seq data. Although these genome annotations were neither complete nor free of errors, which only allowed for approximative evaluations, these were nonetheless useful for assessing mTim's transcript reconstruction accuracy relative to other methods. 

\subsection{Evaluation of transcript reconstruction accuracy}

We evaluated the accuracy of transcripts reconstructed by mTim in a whole-genome comparison to annotated protein-coding genes
using cross-validation (see Methods for details). Here we used two popular criteria that evaluate intron and transcript quality respectively. The first is an assessment of the total number of introns that are inferred correctly (with single-nucleotide precision), whereas the second counts the number of gene loci for which at least one transcript isoform has been reconstructed correctly (all introns predicted correctly). Note that both criteria do not evaluate transcript starts and ends at nucleotide resolution, because  annotations are generally more uncertain for these than for intron boundaries; in transcript evaluation, however, predicted transcript fusion or split predictions will be regarded as errors.

For both criteria we assessed the sensitivity and precision of predicted transcripts. The former is defined as the proportion of annotated introns (or transcripts) which were inferred correctly, whereas the latter is defined as the proportion of inferred introns (or transcripts) which correctly matched an annotated intron (or transcript). The F-score is an aggregate accuracy measure, defined as the harmonic mean of sensitivity and precision: 
$$F = 2 \cdot \frac{\text{sensitivity} \cdot \text{precision}}{\text{sensitivity} + \text{precision}}$$

In initial assessments we verified the effectiveness of mTim's training algorithm and modeling approach. We first evaluated how efficiently the HM-SVM training exploits the available training data. Intron accuracy quickly reached a level where additional training sequences no longer led to substantial improvements: with as little as $80$ training examples an intron accuracy (F-score) of $0.75$ was exceeded, which was only $6.5$\% below the maximum of $0.812$ (Fig.~\ref{fig:train}a). Transcript reconstruction accuracy continued to improve with additional training examples, although with $250$ training sequences transcript accuracy was less than $10$\% below the maximum of $0.373$ (Fig.~\ref{fig:train}a). Second, we assessed the impact of expression-specific submodels (see Fig.~\ref{fig:model} and Methods) on transcript reconstruction accuracy (Fig.~\ref{fig:train}b). While we observed  little effect on intron reconstruction, we confirmed that submodels were valuable for correctly inferring whole transcripts: with five submodels, transcript accuracy increased by $25$\% relative to the simple model without submodels (Fig.~\ref{fig:train}b). Since expression-specific submodels provided an effective means to group exons with similar expression levels into one transcript and terminate it when expression changes dramatically, we used five submodels for all subsequent mTim experiments. Third, we assessed convergence speed of mTim's optimization approach. Results obtained for a training set consisting of $1,000$ sequences 
suggest that after about $80$ iterations, completed in $<2$ CPU hours, prediction accuracy had converged (Fig.~\ref{fig:train}c).

\subsection{Comparison to other transcript reconstruction methods}

\begin{figure}[!tpb]
\vspace*{-1ex}
\centerline{\includegraphics[width=0.8\columnwidth]{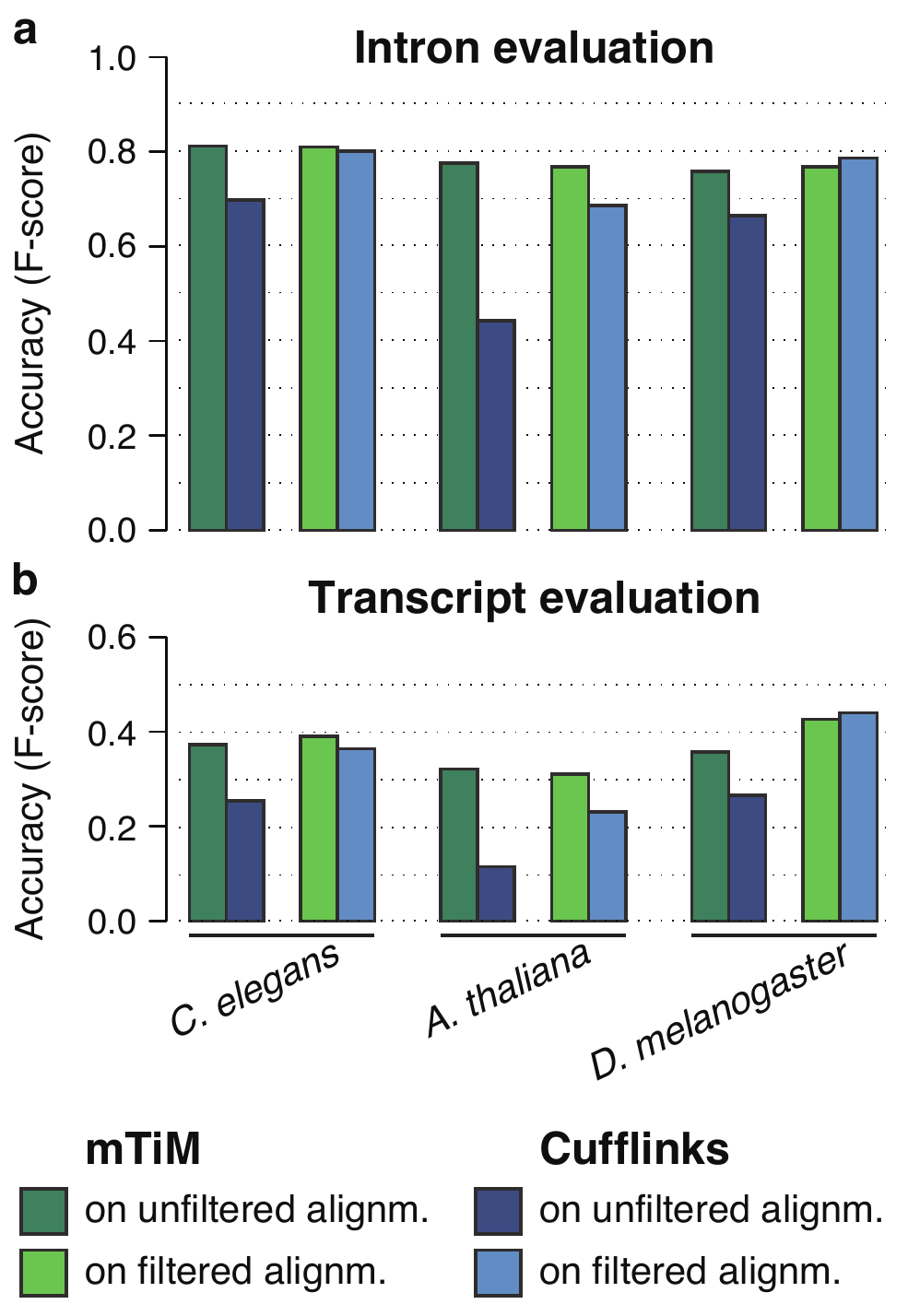}}
\vspace*{-3ex}
\caption{Comparison of transcript reconstruction accuracy between mTim and Cufflinks on RNA-Seq read alignment data from three model organisms. \textbf{(a)} Assessment of the total number of introns whose boundaries were correctly predicted at single-nucleotide precision. \textbf{(b)} Evaluation of the number of gene loci for which at least one transcript isoform was predicted correctly (all introns correct). Methods and whether read alignments were filtered prior to transcript reconstruction are color-coded (see legend). See main text for details and definition of F-score.}\label{fig:compeval}
\end{figure}

To benchmark mTim's transcript reconstruction performance in comparison to other methods, we extended our evaluations to include Cufflinks \cite{CTrapnell2010}, a widely adopted method,
applying the same assessment criteria as before. Comparative evaluations revealed that mTim inferred relatively accurate transcript structures, almost always as good as or better than Cufflinks (Fig.~\ref{fig:compeval}). 
Notably, mTim's predictions were relatively robust against issues in the underlying read alignments (intron accuracy was unaffected by alignment filtering, and transcript F-score decreased by at most $16$\%). Cufflinks in contrast was found to be much more sensitive to these issues; without alignment filtering, its intron and transcript accuracy (F-score) dropped by $13 - 35$\% and $30 - 50$\%, respectively (Fig.~\ref{fig:compeval}). The quality of transcripts inferred by mTim appeared to be relatively high (Fig.~\ref{fig:compeval}) and consistently so across the diverse range of input data tested here; in particular mTim maintained high precision (Table~\ref{tab:compeval}).

\begin{table}[!tpb]
\processtable{Sensitivity and precision of introns and transcripts reconstructed with mTim or Cufflinks applied to PalMapper alignments.\label{tab:compeval}}
{\footnotesize
\begin{tabularx}{\columnwidth}{Xcrrrr}
\hline
\\[-1ex]
& Alignm. & \multicolumn{2}{c}{\bf Sensitivity [\%]} & \multicolumn{2}{c}{\bf Precision [\%]} \\[1ex]
& filtered & mTim & Cufflinks & mTim & Cufflinks \\[1ex]
\hline
\\[-1ex]
& & \multicolumn{4}{c}{\bf Intron evaluation} \\[1ex]
\hline
\\[-1ex]
\textit{C.\ elegans}      & NO & \textbf{75.4} & 58.6 & \textbf{88.1} & 86.4\\
                         & YES & \textbf{74.0} & 71.3 & 89.5 & \textbf{91.6}\\[1ex]
\textit{A.\ thaliana}     & NO & \textbf{69.4} & 30.9 & \textbf{87.8} & 77.9\\
                         & YES & \textbf{69.1} & 53.5 & 86.5 & \textbf{95.6}\\[1ex]
\textit{D.\ melanogaster} & NO & \textbf{70.5} & 66.6 & \textbf{82.1} & 66.5\\
                         & YES & 68.1 & \textbf{70.7} & 88.0 & \textbf{88.6}\\[1ex]
\hline
\\[-1ex]
& & \multicolumn{4}{c}{\bf Transcript evaluation} \\[1ex]
\hline
\\[-1ex]
\textit{C.\ elegans}      & NO & \textbf{30.8} & 20.3 & \textbf{47.3} & 33.8\\
                         & YES & \textbf{31.5} & 30.4 & \textbf{51.2} & 45.2\\[1ex]
\textit{A.\ thaliana}     & NO & \textbf{24.6} &  8.6 & \textbf{46.2} & 17.0\\
                         & YES & \textbf{23.9} & 21.2 & \textbf{44.2} & 25.2\\[1ex]
\textit{D.\ melanogaster} & NO & \textbf{28.0} & 24.7 & \textbf{49.4} & 28.7\\
                         & YES & 32.1 & \textbf{34.7} & \textbf{63.0} & 59.8\\[1ex]
\hline
\end{tabularx}
}{Accuracy values of the best-performing method in each category are in bold face. See main text for definitions of sensitivity and precision and details on alignment filtering.}
\end{table}

\subsection{Flexibility of mTim's approach}
\begin{figure}[!tpb]
\vspace*{-1ex}
\centerline{\includegraphics[width=0.8\columnwidth]{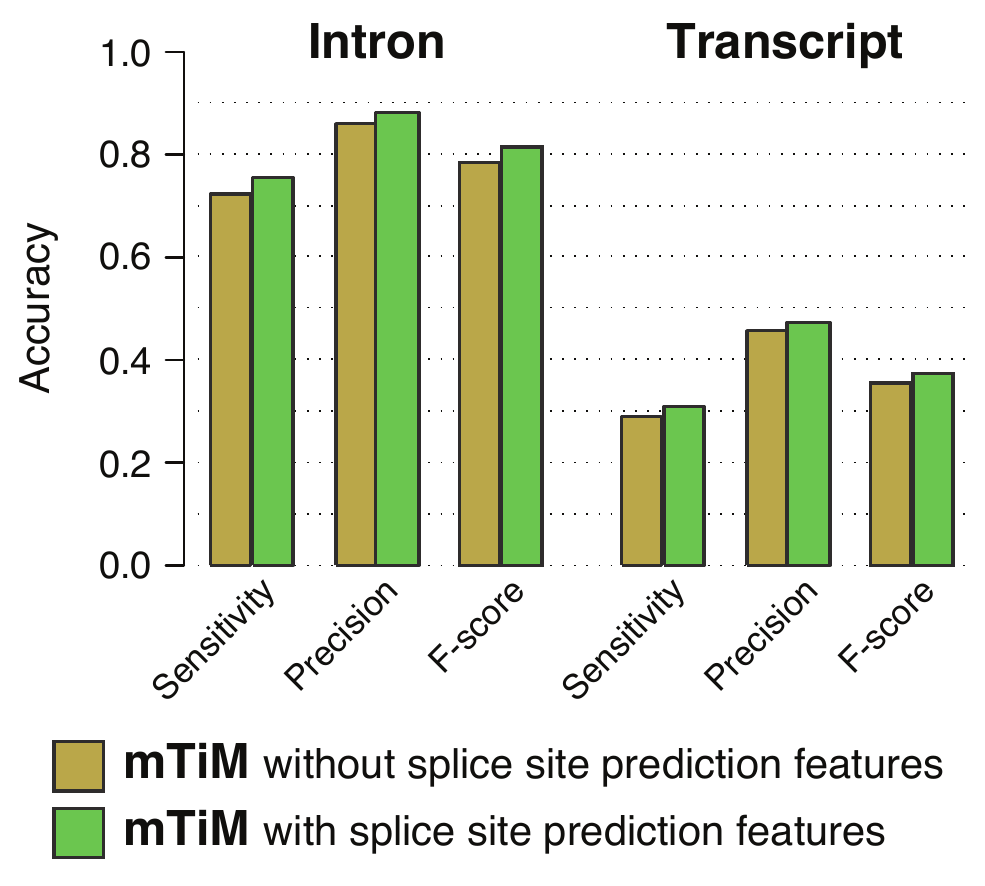}}
\vspace*{-3ex}
\caption{Accuracy of mTim when trained with and without features derived from genomic sequence signals around splice sites (see Methods for details). Both mTim instances were trained and evaluated on unfiltered RNA-Seq alignments from \textit{C.\ elegans}.}\label{fig:splicesites}
\end{figure}

Due to its modular architecture and its general machine-learning approach, mTim can easily be tailored to specific application requirements. For instance features corresponding to genomic splice site predictions can be disabled, making mTim rely completely on RNA-Seq alignment features thereby eliminating any potential bias against non-coding transcripts. We assessed the extent to which this affects transcript reconstruction accuracy and found the effect to be minor (Fig.~\ref{fig:splicesites}). 

Extensions of mTim's feature set are easily possible as well. Future developments could include additional features derived from promoter predictions \cite{SSonnenburg2006Bioinf}, transcription factor ChIP-Seq data or methylation experiments (see e.g.\ \cite{MGerstein2010}), all of which might be useful to better recognize transcript start and end sites, which is a common source of errors with the current approach.

\section{Conclusion}

Here, we have introduced mTim, a discriminative machine learning-based method that reconstructs transcripts from RNA-Seq read alignments and splice site predictions. 
We have shown that it is able to infer transcripts with high accuracy and that it is more robust errors in the underlying read alignments.
Pre-trained mTim predictors used for this work are available within the Oqtans Galaxy webserver (\url{http://oqtans.org/}). Moreover, mTim is open-source software provided via \url{https://github.com/nicococo/mTIM}.

\section*{Acknowledgement}

We are grateful to Klaus-Robert M{\"u}ller, Christian Widmer, Marius Kloft, and Vipin Sreedharan for insightful comments and discussions, further to Andre Noll for technical support.
\paragraph{Funding\textcolon} This work was supported by the Max  Planck Society (GR, PM), by the German Research Foundation (NG, SS, grants DFG MU 987/6-1, RA 1894/1-1), and an EMBL postdoctoral fellowship (GZ).
We would also like to acknowledge support through the BMBF project "ALICE, Autonomous Learning in Complex Environments" (01IB10003B).

%
\bibliographystyle{plain}
\bibliography{paper}

\end{document}